\documentclass[12pt,twoside]{article}
\usepackage{a4}

\begin{document}

\newcommand{\gf}{G_{\mbox{{\scriptsize F}}}}
\newcommand{\order}{{\cal O}}
\newcommand{\delp}{\Delta P}
\newcommand{\leqsim}{\,\mbox{{\scriptsize $\stackrel{<}{\sim}$}}\,}
\newcommand{\geqsim}{\,\mbox{{\scriptsize $\stackrel{>}{\sim}$}}\,}
\newcommand{\bskk}{B_s\to K^+ K^-}
\newcommand{\bsokk}{B_s^0\to K^+ K^-}
\newcommand{\bsobkk}{\bar B_s^0\to K^+ K^-}
\newcommand{\bppipko}{B^+\to \pi^+ K^0}
\newcommand{\bmpimkob}{B^-\to \pi^- \bar K^0}
\newcommand{\bmpiokm}{B^-\to \pi^0 K^-}
\newcommand{\bppiokp}{B^+\to \pi^0 K^+}
\newcommand{\bdpimkp}{B^0_d\to \pi^- K^+}
\newcommand{\bdbpipkm}{\bar B^0_d\to \pi^+ K^-}
\newcommand{\alphaqed}{\alpha_{\mbox{{\scriptsize QED}}}}
\newcommand{\pew}{P_{\mbox{{\scriptsize EW}}}}
\newcommand{\pewb}{\bar P_{\mbox{{\scriptsize EW}}}}
\newcommand{\pewp}{P_{\mbox{{\scriptsize EW}}}'}
\newcommand{\pewpb}{\bar P_{\mbox{{\scriptsize EW}}}'}
\newcommand{\pewc}{P_{\mbox{{\scriptsize EW}}}^{\mbox{{\scriptsize C}}}}
\newcommand{\pcew}{P_{\mbox{{\scriptsize EW}}}^{\mbox{{\scriptsize (C)}}}}
\newcommand{\pcewb}{\bar P_{\mbox{{\scriptsize
EW}}}^{\mbox{{\scriptsize (C)}}}}
\newcommand{\pewpc}{P_{\mbox{{\scriptsize EW}}}'^{\mbox{{\scriptsize C}}}}
\newcommand{\bbtodb}{\bar b\to\bar d}
\newcommand{\bbtosb}{\bar b\to\bar s}

\newcommand{\bkk}{B_d\to K^0\bar K^0}
\newcommand{\bksks}{B_d\to K_{\mbox{{\scriptsize S}}}
K_{\mbox{{\scriptsize S}}}}
\newcommand{\bpiphi}{B_s\to\pi^0\Phi}
\newcommand{\bopiphi}{B_s^0\to\pi^0\Phi}
\newcommand{\bbopiphi}{\bar B_s^0\to\pi^0\Phi}
\newcommand{\brhok}{B_s\to\rho^0K_{\mbox{{\scriptsize S}}}}
\newcommand{\bqf}{B_q\to f}
\newcommand{\pcps}{\phi_{\mbox{{\scriptsize CP}}}(B_s)}
\newcommand{\pcpq}{\phi_{\mbox{{\scriptsize CP}}}(B_q)}
\newcommand{\pw}{\phi_{\mbox{{\scriptsize W}}}}
\newcommand{\acp}{a_{\mbox{{\scriptsize CP}}}}
\newcommand{\acpdir}{{\cal A}_{\mbox{{\scriptsize CP}}}^
{\mbox{{\scriptsize dir}}}}
\newcommand{\acpmi}{{\cal A}_{\mbox{{\scriptsize
CP}}}^{\mbox{{\scriptsize mix-ind}}}}
\newcommand{\acc}{A_{\mbox{{\scriptsize CC}}}}
\newcommand{\aew}{A_{\mbox{{\scriptsize EWP}}}}
\newcommand{\heff}{{\cal H}_{\mbox{{\scriptsize eff}}}(\Delta B=-1)}
\newcommand{\hefftot}{{\cal H}_{\mbox{{\scriptsize eff}}}(|\Delta B|=1)}
\newcommand{\heffp}{{\cal H}_{\mbox{{\scriptsize eff}}}(\Delta B=+1)}
\newcommand{\xif}{\xi_f^{(q)}}
\newcommand{\xiskk}{\xi_{K^+K^-}^{(s)}}
\newcommand{\xikk}{\xi_{K^0\bar K^0}^{(d)}}
\newcommand{\VmA}{\mbox{{\scriptsize V--A}}}
\newcommand{\VpA}{\mbox{{\scriptsize V+A}}}
\newcommand{\VpmA}{\mbox{{\scriptsize V$\pm$A}}}
\newcommand{\beq}{\begin{equation}}
\newcommand{\eeq}{\end{equation}}
\newcommand{\bea}{\begin{eqnarray}}
\newcommand{\eea}{\end{eqnarray}}
\newcommand{\non}{\nonumber}
\newcommand{\lab}{\label}
\newcommand{\la}{\langle}
\newcommand{\ra}{\rangle}
\newcommand{\np}{Nucl.\ Phys.}
\newcommand{\pl}{Phys.\ Lett.}
\newcommand{\prl}{Phys.\ Rev.\ Lett.}
\newcommand{\pr}{Phys.\ Rev.}
\newcommand{\zp}{Z.\ Phys.}

\setcounter{page}{-1}
\thispagestyle{empty}
\begin{flushright}
TTP95-32\\
\vspace{0.2cm}
hep-ph/9509204\\
\vspace{0.2cm}
August 1995
\end{flushright}
\vspace{2cm}
\begin{center}
{\Large{\bf Strategies for Fixing the CKM-angle $\gamma$ and}}\\
\vspace{0.3cm}
{\Large{\bf Obtaining Experimental Insights into the}}\\
\vspace{0.3cm}
{\Large{\bf World of Electroweak Penguins}}\\
\vspace{2.3cm}

{\large{\sc Robert Fleischer}}\\
\vspace{0.3cm}
{\sl Institut f\"ur Theoretische Teilchenphysik\\
\vspace{0.1cm}
Universit\"at Karlsruhe\\
\vspace{0.1cm}
D--76128 Karlsruhe, Germany}\\
\vspace{1.8cm}
{\large{\bf Abstract}}
\end{center}
\vspace{0.7cm}
Using the $SU(3)$ flavour symmetry of strong interactions, we propose
strategies for extracting both the CKM-angle $\gamma$ and the
$\bar b\to\bar uu\bar s$ tree-level amplitude $T'$. We present also
an approximate approach using the branching ratios for the modes
$B^+\to\pi^+ K^0$, $B^0_d\to\pi^- K^+$, $\bar B^0_d\to\pi^+ K^-$ and
$B^+\to\pi^+\pi^0$ which should be rather promising from the experimental
point of view. The quantities $\gamma$ and $T'$ determined this way may well
be used as an input to control electroweak penguins in nonleptonic $B$-decays
as has been discussed in previous work. Following these lines, we propose
strategies for obtaining quantitative insights into the physics of the
electroweak penguin operators and performing some consistency checks. As a
by-product, we derive an upper bound of $6^\circ$ for the uncertainty
originating from electroweak penguins in the $\alpha$-determination by means
of $B\to\pi\pi$ decays.
\newpage
\thispagestyle{empty}
\mbox{}
\newpage
Strategies for the determination of the angle $\gamma$ in the unitarity
triangle \cite{ck,js} are among the central issues of present particle
physics phenomenology. Although there are already methods on the market
allowing an absolutely clean measurement of this quantity (see e.g.\
refs.~\cite{gw}-\cite{kay}), they are quite challenging for
experimentalists. An interesting approach to measure both weak and strong
phases by using $SU(3)$ triangle relations among $B\to\{\pi\pi,\pi K,
K\bar K\}$ decays and making some plausible dynamical assumptions (neglect
of annihilation topologies, etc.) was proposed last year by Gronau,
Hern\'andez, London and Rosner \cite{grl}-\cite{ghlrsu3}.

Unfortunately, similar to the situation arising in certain nonleptonic
$B$-meson decays \cite{rfewp1}-\cite{dy}, electroweak penguins may have a
considerable impact on this approach and may in particular preclude a clean
determination of the CKM-angle $\gamma$ \cite{dhewp2,ghlrewp}. In order to
eliminate the electroweak penguin contributions, Gronau et al.\ have
constructed an amplitude quadrangle involving $B\to\pi K$ decays
\cite{ghlrewp} that
can be used in principle to extract $\gamma$. However, this approach is
very difficult from the experimental point of view, since one diagonal of
the quadrangle corresponds to the decay $B_s\to\pi^0\eta$ which is
expected to have a very small branching ratio at the $\order(10^{-7})$
level. Recently, Deshpande and He have presented another $SU(3)$-based
method \cite{dhgam} which uses the charged $B$-decays
$B^-\to\{\pi^-\bar K^0,\pi^0K^-,\eta K^-,\pi^-\pi^0\}$ and is
unaffected by electroweak penguins as well. Although this approach is more
promising for experimentalists -- the relevant branching ratios are
$\order(10^{-5})$ -- it suffers from $\eta-\eta'-$mixing and other
$SU(3)$-breaking effects.

In a recent publication \cite{bfewp}, it has been shown that the $\bbtosb$
electroweak penguin amplitude $(c_u-c_d)\pewp$ can be determined
from $B\to\pi K$ decays having branching ratios
$\order(10^{-5})$ if one uses the CKM-angle $\gamma$ as one
of the central inputs and makes some reasonable approximations. Since
electroweak penguins are -- in contrast to QCD penguins \cite{bf} --
dominated to a good approximation by internal top-quark exchanges, the
$\bbtosb$ electroweak penguin amplitude can be related to the $\bbtodb$
amplitude \mbox{$(c_u-c_d)\pew$} by using the $SU(3)$ flavour symmetry of
strong interactions. The knowledge of $(c_u-c_d)\pew$ allows in particular
to investigate whether the uncertainty $\Delta\alpha$
in the Gronau--London--method \cite{gl}
of measuring the CKM-angle $\alpha$ introduced by electroweak
penguin effects is really as small as is expected from theoretical
estimates \cite{dhewp2,ghlrewp}. As we shall see below, an upper limit
for this uncertainty is given by $|\Delta\alpha|\leqsim6^\circ$.

In this letter we would like to discuss some other applications of the
approach \cite{bfewp} which allow {\it quantitative} insights into the
physics of the electroweak penguin operators and provide interesting
tests of the Standard Model of electroweak interactions \cite{sm}.
As the CKM-angle $\gamma$
is one of the central ingredients of this method, let us begin our
discussion by presenting a new strategy for extracting this quantity. To
this end we consider the decays $\bppipko$ and $\bdpimkp$. If we apply
the same notation as Gronau, Hern\'andez, London and Rosner in ref.\
\cite{ghlrewp}, the corresponding decay amplitudes take the form
\beq\lab{e1}
\begin{array}{rcl}
A(\bppipko)&=&P'+c_d\pewpc\\
A(\bdpimkp)&=&-(P'+T'+c_u\pewpc),
\end{array}
\eeq
where $T'$ describes the colour-allowed $\bar b\to\bar uu\bar s$ tree-level
amplitude, $P'$ denotes $\bbtosb$ QCD penguins and $\pewpc$ is related
to colour-suppressed electroweak penguins. Neglecting as in \cite{bfewp}
the colour-suppressed electroweak penguin contributions, we obtain
\beq\lab{e2}
\begin{array}{rcl}
A(\bppipko)&=&P'\,\,=\,\, A(\bmpimkob)\\
A(\bdpimkp)&=&-(P'+T')\\
A(\bdbpipkm)&=&-(P'+e^{-2i\gamma}T'),
\end{array}
\eeq
where we have taken into account the relations
\beq\lab{e3}
P'=|P'|e^{i\delta_{P'}}e^{i\pi}=\bar P'
\eeq
\beq\lab{e4}
\bar T'=e^{-2i\gamma}T'.
\eeq
Note that (\ref{e3}) arises from the special CKM-structure of the $\bbtosb$
penguins \cite{bf} and that $\delta_{P'}$ and $\pi$ are CP-conserving
strong and CP-violating weak phases, respectively. In Fig.~1 we have
represented (\ref{e2}) graphically in the complex plane.
Looking at this figure, which is a modification of Fig.~2 given in
ref.~\cite{bfewp}, implies that both the quantities $z\equiv T'/|P'|$,
$\bar z\equiv\bar T'/|P'|$ and the CKM-angle $\gamma$
can be determined from the measured branching
ratios BR$(\bppipko)=\mbox{BR}(\bmpimkob)$, BR$(\bdpimkp)$ and
BR$(\bdbpipkm)$, if $|T'|=|\bar T'|$ is known. This quantity can be
determined by relating it to the $\bar b\to\bar uu\bar d$ colour-allowed
tree-level amplitude $T$. Making use of the $SU(3)$  flavour symmetry of
strong interactions, one finds \cite{ghlrsu3}
\beq\lab{e5}
\frac{|T'|}{|T|}=\lambda\frac{f_K}{f_{\pi}},
\eeq
where $\lambda=0.22$ is the usual Wolfenstein parameter \cite{wolf} and
$f_K$ and $f_\pi$ are the $K$- and $\pi$-meson decay constants,
respectively, describing {\it factorizable} $SU(3)$-breaking. Unfortunately,
non-factorizable $SU(3)$-breaking corrections to (\ref{e5}) are completely
unknown at present and therefore cannot be included. In view of these
uncertainties, the neglect of the colour-suppressed electroweak penguins
in (\ref{e2}), which are expected to be suppressed by factors
$\order(\bar\lambda^2)$ and $\order(\bar\lambda)$ relative to the
amplitudes $P'$ and $T'$, respectively, seems to be reasonable. The
parameter $\bar\lambda=\order(0.2)$ describes the hierarchy of the different
topologies contributing to $B\to PP$ decays \cite{ghlrsu3,ghlrewp,bfewp}.

The quantity $|T|$ can be extracted in principle in a clean way
up to corrections of $\order(\bar\lambda^2)$ by applying
the approach presented in \cite{bfalpha}.  There, a determination
of the CKM-angle $\alpha$ by using mixing-induced CP violation in the
decays $B_d\to\pi^+\pi^-$ and $B_d\to K^0\bar K^0$ has been proposed. As
a by-product of this analysis, the quantity $|T+E|$, where $E$ corresponds
to an $\order(\bar\lambda^2)$-suppressed exchange amplitude, is also fixed
and can be used to determine $|T'|$ through (\ref{e5}). This approach is,
however, quite difficult for experimentalists. Therefore, it is an important
question to search also for methods of obtaining {\it approximate}
information on $|T|$ that can be realized more easily in practice.

Such an estimate of $|T|$ can be obtained directly from the branching
ratio for the decay $B^+\to\pi^+\pi^0$. In the notation
of ref.~\cite{ghlrewp} its transition amplitude takes the form
\beq\lab{e6}
A(B^+\to\pi^+\pi^0)=-\frac{1}{\sqrt{2}}\left[T+C+(c_u-c_d)(\pew+\pewc)\right].
\eeq
If we neglect both the colour-suppressed $\bar b\to\bar uu\bar d$ tree-level
amplitude $C$ and the very small electroweak penguin contributions, which
should be suppressed relative to $T$ by factors $\order(\bar\lambda)$ and
$\order(\bar\lambda^2)$, respectively \cite{ghlrewp}, we find
\beq\lab{e7}
|T|\approx\sqrt{2}|A(B^+\to\pi^+\pi^0)|.
\eeq
This approximation is consistent with those performed in \cite{bfewp} where
the colour-suppressed $\bar b\to\bar uu\bar s$ tree-level amplitude $|C'|$ has
also been neglected. Consequently, combining (\ref{e5}) and (\ref{e7}), the
amplitudes $z$ and $\bar z$, which are essential for the determination of
the $\bbtosb$ electroweak penguin amplitude $(c_u-c_d)\pewp$ as has been
outlined in \cite{bfewp}, can be extracted from Fig.~1. Since the angle
between these amplitudes is given by $2\gamma$ it can be fixed as well. This
determination of $\gamma$ is just an {\it estimate} and cannot be considered
as a precision measurement. However, it should be rather promising from the
experimental point of view because all involved branching ratios are
$\order(10^{-5})$. Moreover, it requires a quite simple geometrical
construction, which is very similar to the original one suggested by
Gronau, Rosner and London \cite{grl}, and is not affected by
$\eta-\eta'-$mixing as the method for determining $\gamma$ proposed by
Deshpande and He \cite{dhgam}.

The quantities $z$ and $\bar z$ may well be used to obtain information on
$\bbtosb$ electroweak penguins by following the approach presented in
ref.~\cite{bfewp}. In this paper, different strategies for determining $z$
and $\bar z$ that can be used if the CKM-angle $\gamma$ is known, for
example by applying the absolutely clean method of Gronau and Wyler
\cite{gw}, have also been discussed.

The scenario at future experimental $B$-physics facilities might be as
follows:
\begin{itemize}
\item In the $1^{\mbox{{\scriptsize st}}}$ generation of experiments
it should be possible to estimate $z$, $\bar z$ and $\gamma$ by measuring
the four branching ratios BR$(\bppipko)=\mbox{BR}(\bmpimkob)$, BR$(\bdpimkp)$,
BR$(\bdbpipkm)$ and BR$(B^+\to\pi^+\pi^0)$ as we have proposed above.
If one measures in addition the two branching ratios BR$(\bppiokp)$
and BR$(B^-\to \pi^0 K^-)$, the $\bbtosb$ electroweak penguin amplitude
$(c_u-c_d)\pewp$ can be determined (see Fig.~1 given in \cite{bfewp}).
Note that all involved branching ratios are expected to be at the
$\order(10^{-5})$ level \cite{kp} and that no time-dependent measurements
are needed.
\item In the $2^{\mbox{{\scriptsize nd}}}$ --
or probably $3^{\mbox{{\scriptsize rd}}}$ -- generation of experiments the
CKM-angle $\gamma$ can hopefully be determined by using the absolutely clean
methods of \cite{gw}-\cite{adk}. The knowledge of this angle would improve the
determination of the electroweak penguin amplitude $(c_u-c_d)\pewp$
considerably as has been stressed in \cite{bfewp}. In these experiments it
will hopefully also be possible to implement the approaches \cite{gl,bfalpha}
to determine the CKM-angle $\alpha$ in a clean way. The uncertainties of
both methods, electroweak penguins and $SU(3)$-breaking in \cite{gl} and
\cite{bfalpha}, respectively, are expected to be of the same order,
i.e.\ $\order(\bar\lambda^2)$, and should be rather small. In the case of
the $\alpha$-determination by means of $B\to\pi\pi$ decays \cite{gl}, the
electroweak penguin corrections can even be controlled in a quantitative
way as has been shown in \cite{bfewp}. Besides the extraction of $\alpha$,
the approach presented in \cite{bfalpha} would also allow an independent
determination of $\gamma$ and $z$, $\bar z$ as we have shown above.
This would yield an interesting cross-check.
\end{itemize}

Let us now discuss some further applications of the approach to extract the
electroweak penguin contributions to nonleptonic $B$-decays presented
in \cite{bfewp}. They can be divided into two categories:
\begin{itemize}
\item[i)]{\it quantitative insights into the physics of the electroweak
penguin operators}
\item[ii)]{\it consistency checks}.
\end{itemize}
In order to work out point i), we shall use a low energy effective
Hamiltonian describing $|\Delta B|=1$, $\Delta C=\Delta U=0$
transitions. In the case of $b\to s$ decays, the $\Delta B=-1$ part can
be written in the form \cite{bjl}
\bea
\lefteqn{\heff=}\lab{e8}\\
&&\frac{\gf}{\sqrt{2}}\left[V_{us}^\ast V_{ub}\sum\limits_{k=1}^2Q_k^u
C_k(\mu)+V_{cs}^\ast V_{cb}\sum\limits_{k=1}^2Q_k^c C_k(\mu)
-V_{ts}^\ast V_{tb}\sum\limits_{k=3}^{10}Q_k C_k(\mu)\right],\nonumber
\eea
where $Q_k$ are local four-quark operators and $C_k(\mu)$ denote
so-called Wilson coefficient functions that have been calculated by Buras
et al.\ \cite{bjl} in renormalization group improved perturbation theory
including both leading and next-to-leading order QCD corrections and
leading order corrections in the QED coupling $\alphaqed$. As usual,
$\mu$ denotes a renormalization scale $\order(m_b)$. For details
concerning phenomenological applications of the next-to-leading order
Hamiltonian (\ref{e8}) to nonleptonic $B$-decays, the reader is referred to
refs.~\cite{rfewp1,rfewp2,rf,kps} where also the exact definitions of the
current-current operators $Q_{1/2}^u$, $Q_{1/2}^c$, the QCD penguin
operators $Q_3,\ldots,Q_6$ and the electroweak penguin operators
$Q_7,\ldots,Q_{10}$ can be found.

The aim of the following discussion is to derive a transparent expression
that allows a quantitative test whether the electroweak
penguin amplitude $(c_u-c_d)\pewp$ determined by following the approach
\cite{bfewp} is consistent with the description through the
Standard Model of electroweak interactions~\cite{sm}.
To this end let us neglect the QCD corrections to the
electroweak penguin operators. In view of the approximations that will be
made in a moment, this approximation seems to be reasonable. Moreover,
these QCD corrections should be small if one defines the top-quark mass
properly as $\overline{m_t}(m_t)$ \cite{bb} (see also ref.~\cite{rfewp3}).
Within this approximation, the Wilson coefficients of the electroweak
penguin operators are given by the functions $\bar C^{(0)}_k(\mu)$
specified in refs.~\cite{rfewp1,rfewp2}. Since $\bar C^{(0)}_8(\mu)$ and
$\bar C^{(0)}_{10}(\mu)$ vanish, we have to consider only the hadronic
matrix elements of the electroweak penguin operators $Q_7$ and $Q_9$. Note
that the operator $Q_9$ plays the most important role because of its large
Wilson coefficient \cite{rfewp1}.

The modes that are used in \cite{bfewp} to determine the $\bbtosb$
electroweak penguins are $\bppiokp$ and $\bmpiokm$. The corresponding
$\Delta B=-1$ hadronic matrix elements of the electroweak penguin
operators $Q_7$ and $Q_9$ are given by
\bea
\langle\pi^0K^-|Q_7|B^-\rangle&=&\left\langle\pi^0K^-\left|\left\{(\bar
uu)_{\VpA}-\frac{1}{2}(\bar dd)_{\VpA}\right\}(\bar sb)_{\VmA}\right|B^-
\right\rangle
\lab{e10a}\\
\langle\pi^0K^-|Q_9|B^-\rangle&=&\left\langle\pi^0K^-\left|\left\{(\bar
uu)_{\VmA}-\frac{1}{2}(\bar dd)_{\VmA}\right\}(\bar sb)_{\VmA}
\right|B^-\right\rangle.
\lab{e10b}
\eea
In the notation of Gronau et al.~\cite{ghlrewp},
the matrix elements (\ref{e10a}) and (\ref{e10b})
are incorporated in the electroweak penguin amplitude $(c_u-c_d)\pewpb$.
On the other hand, the colour-allowed amplitude $\bar T'$ is related to
hadronic matrix elements of the current-current operators $Q_1^u$ and $Q_2^u$.
In order to calculate this quantity, we consider the mode $\bdbpipkm$
since -- in contrast to $\bmpiokm$ -- no colour-suppressed amplitude $\bar C'$
is present in this case. The corresponding hadronic matrix elements of the
current-current operators $Q_1^u$ and $Q_2^u$ are given by
\bea
\langle\pi^+K^-|Q_1^u|\bar B^0_d\rangle&=&\langle\pi^+K^-|(\bar s_{\alpha}
u_{\beta})_{\VmA}(\bar u_{\beta}b_{\alpha})_{\VmA}|\bar B^0_d
\rangle\lab{e11a}\\
\langle\pi^+K^-|Q_2^u|\bar B^0_d\rangle&=&\langle\pi^+K^-|(\bar su)_{\VmA}
(\bar ub)_{\VmA}|\bar B^0_d\rangle,\lab{e11b}
\eea
where $\alpha$ and $\beta$ denote $SU(3)_{\mbox{\scriptsize C}}$ colour
indices. Note that QCD penguin matrix elements of the current-current
operators $Q_2^u$ and $Q_2^c$ \cite{rf}, where the up- and charm-quarks
run as virtual particles in the loops, respectively, contribute
{\it by definition} to the QCD penguin amplitude $\bar P'$ and {\it not}
to $\bar T'$. A similar comment applies also to the effects of inelastic
final state interactions \cite{kam} that originate e.g.\ from the
rescattering process $\bar B^0_d\to\{D^-_sD^+\}\to\pi^+K^-$. In our notation,
these contributions are related to penguin-like matrix elements of the
current-current operators and are also included in $\bar P'$.

If we introduce non-perturbative $B$-parameters and apply the
$SU(3)$ flavour symmetry of strong interactions, i.e.\ do not
distinguish between $u$, $d$ and $s$ quark-flavours, we obtain
\beq\lab{e11c}
\begin{array}{rclrcl}
\langle\pi^0K^-|Q_7(\mu)|B^-\rangle&=&\frac{1}{\sqrt{2}}\frac{3}{2}
\tilde B_2(\mu)f,&
\langle\pi^0K^-|Q_9(\mu)|B^-\rangle&=&\frac{1}{\sqrt{2}}\frac{3}{2}
B_2(\mu)f,\\
\langle\pi^+K^-|Q_1^u(\mu)|\bar B^0_d\rangle&=&\frac{1}{3}B_1(\mu)f,&
\langle\pi^+K^-|Q_2^u(\mu)|\bar B^0_d\rangle&=&B_2(\mu)f.
\end{array}
\eeq
In these equations, the quantity $f$ corresponds to the ``factorized''
matrix element $\langle K^-|(\bar su)_{\VmA}|0\rangle\langle\pi^+|
(\bar ub)_{\VmA}|\bar B^0_d\rangle$. It is quite natural to assume
$\tilde B_2(\mu)\approx -B_2(\mu)$, since
the $\pi^0$-meson is a pseudoscalar particle and therefore emerges from
the axial-vector parts of the quark-currents $[(\bar uu)_{\VpmA}-
(\bar dd)_{\VpmA}]$ arising in the electroweak penguin operators $Q_7$
and $Q_9$. For a similar reason, the one-loop QED penguin matrix elements
of the current-current operators $Q_{1/2}^u$ and $Q_{1/2}^c$ vanish
and do not contribute to the amplitude \mbox{$(c_u-c_d)\pewpb$}. The point
is that the virtual photons appearing in the QED penguin diagrams
generate $(\bar uu)_{\mbox{{\scriptsize V}}}$ and
$(\bar dd)_{\mbox{{\scriptsize V}}}$ vector-currents that cannot create the
pseudoscalar $\pi^0$-meson \cite{rfewp3}.

Combining all these considerations, we eventually arrive at
\beq\lab{e12b}
(c_u-c_d)\pewpb=\frac{\gf}{\sqrt{2}}V_{ts}^\ast V_{tb}\frac{3}{2}\left[\bar
C^{(0)}_9(\mu)-\bar C^{(0)}_7(\mu)\right]B_2(\mu)f
\eeq
\beq\lab{e12a}
\bar T'=-\frac{\gf}{\sqrt{2}}V_{us}^\ast V_{ub}\left[
\frac{1}{3}\frac{B_1(\mu)}{B_2(\mu)}C_1(\mu)+C_2(\mu)\right]B_2(\mu)f.
\eeq
Factorizable $SU(3)$-breaking affecting (\ref{e12b}) can be taken into
account approximately by multiplying its r.h.s.\ by
\beq\lab{e15b}
r_{SU(3)}\equiv\frac{f_{\pi}}{f_K}\frac{F_{BK}(0;0^+)}{F_{B\pi}(0;0^+)}.
\eeq
Unfortunately, this correction factor depends not only on pseudoscalar
meson decay constants as (\ref{e5}), but also on model-dependent form
factors $F_{BK}(0;0^+)$, $F_{B\pi}(0;0^+)$ parametrizing hadronic
quark-current matrix elements \cite{bgr}. The model of Bauer, Stech
and Wirbel \cite{bsw} yields $r_{SU(3)}\approx1$ indicating that
factorizable $SU(3)$-breaking to (\ref{e12b}) is small. At present
non-factorizable $SU(3)$-breaking is completely unknown. Consequently,
we cannot include these corrections and obtain
\beq\lab{e14}
{\cal R}_{\mbox{{\scriptsize EW}}}\equiv
\frac{|(c_u-c_d)\pewp|}{|T'|}\approx\frac{3}{2}\left\vert\frac{V_{ts}^\ast
V_{tb}}{V_{us}^\ast V_{ub}}\right\vert\cdot
\left\vert\frac{\bar C^{(0)}_9(\mu)-\bar C^{(0)}_7(\mu)}{\frac{1}{3}
\frac{B_1(\mu)}{B_2(\mu)}C_1(\mu)+C_2(\mu)}\right\vert r_{SU(3)}.
\eeq
Applying the Wolfenstein expansion of the CKM-matrix \cite{wolf} yields
\beq\lab{e14o}
\left\vert\frac{V_{ts}^\ast V_{tb}}{V_{us}^\ast V_{ub}}\right\vert=\frac{1}
{\lambda^2R_b}\left(1+\order(\lambda^2)\right)\approx\frac{1}{\lambda^2R_b},
\eeq
where the parameter $R_b\equiv|V_{ub}|/(\lambda|V_{cb}|)$
is constrained by present experimental data to lie within the range
$R_b=0.36\pm0.08$ \mbox{\cite{blo,al}}.
In order to eliminate the combination of the Wilson coefficients
$C_{1/2}(\mu)$ and the non-perturbative $B$-parameters $B_{1/2}(\mu)$
appearing in the denominator of eq.~(\ref{e14}), we identify
it with the phenomenological parameter $a_1$ \cite{bsw,nrsx,cleo,bur}.
Present experimental data implies $a_1\approx1.05\pm0.10$.
Applying the analytical expressions for the Wilson coefficients
$\bar C^{(0)}_9(\mu)$
and $\bar C^{(0)}_7(\mu)$ given in refs.~\cite{rfewp1,rfewp2},
the $\mu$-dependences of these coefficients cancel explicitly and we get
the $\mu$-independent result
\beq\lab{e15}
{\cal R}_{\mbox{{\scriptsize EW}}}\approx
\frac{\alphaqed}{2\pi\lambda^2R_ba_1\sin^2\Theta_{\mbox{{\scriptsize
W}}}}\left|5B(x_t)-2C(x_t)\right|r_{SU(3)},
\eeq
where $x_t\equiv m_t^2/M_W^2$ introduces a top-quark mass dependence
into this expression and $B(x_t)$ and $C(x_t)$ are two of the well-known
Inami--Lim functions \cite{il}. In Fig.~2 we have shown the
expected dependence of ${\cal R}_{\mbox{{\scriptsize EW}}}$ described by
eq.~(\ref{e15}) on the top-quark mass $m_t$ for various values of the
CKM-parameter $R_b$. In drawing this figure, we have used
$a_1=1$ and $r_{SU(3)}=1$. Note that the results for
${\cal R}_{\mbox{{\scriptsize EW}}}$ shown in
Fig.~2 are remarkably consistent with the na\"\i ve hierarchy given by
Gronau et al.\ in \cite{ghlrsu3,ghlrewp}
yielding ${\cal R}_{\mbox{{\scriptsize EW}}}=\order(1)$.

As a by-product, applying the results of ref.~\cite{bfewp},
we can easily estimate an upper bound for the uncertainty $\Delta\alpha$
originating from electroweak penguins in the $\alpha$-determination by
means of the mixing-induced CP asymmetry $\acpmi(B_d\to\pi^+\pi^-)$ and
isospin relations among $B(\bar B)\to\pi\pi$ decay
amplitudes~\mbox{\cite{bfewp,gl}}:
\beq\lab{e16o}
|\Delta\alpha|\leqsim\frac{\alphaqed}{2\pi a_1\sin^2\Theta_{\mbox{{\scriptsize
W}}}}\left|5B(x_t)-2C(x_t)\right|\cdot\left\vert\frac{V_{td}}{V_{ub}}
\right\vert\left\vert\sin\alpha\right\vert.
\eeq
Note that the $SU(3)$-breaking factor $r_{SU(3)}$ cancels in this
expression. Taking into account $|V_{td}|/|V_{ub}|\leq5.8$ \cite{blo,al} and
$|\sin\alpha|\leq1$, we obtain from Fig.~2 $|\Delta\alpha|\leqsim6^\circ$.

\vspace{0.3cm}

Let us now finally come to point ii) listed above discussing some
interesting {\it consistency checks}:
\begin{itemize}

\item{The transitions $B_s\to\pi^0(\eta,\Phi)$:}

The ratio ${\cal R}_{\mbox{{\scriptsize EW}}}$ determined by following
the approach \cite{bfewp} can be used to extract the parameter $x$ that
has been introduced in \cite{rfewp3} to describe the decay
$B_s\to\pi^0\Phi$ through $x\approx-a_1{\cal R}_{\mbox{{\scriptsize EW}}}/
(a_2r_{SU(3)})$. Here, $a_2$ is the so-called phenomenological
colour-suppression factor
\cite{bsw,nrsx,cleo,bur}. A detailed phenomenological analysis of the decay
$B_s\to\pi^0\Phi$ has been performed in \cite{rfewp3}. The dominance
of the electroweak penguins arising in this mode, which has first been
pointed out there, has been confirmed independently by the authors of
refs.\ \mbox{\cite{dht,dy}}. Note that the structure of the decay
$B_s\to\pi^0\Phi$ is very similar to that of the transition
$B_s\to\pi^0\eta$ describing one diagonal of the quadrangle constructed
in \cite{ghlrewp}. Unfortunately, the branching ratios for the
decays $B_s\to\pi^0(\eta,\Phi)$ are very small and are
expected to be of $\order(10^{-7})$. Following the strategy
presented above, it should be possible to predict them and the
corresponding CP asymmetries on a rather solid ground {\it before}
one can measure these quantities. It should also be possible to predict
the shape of the quadrangle derived in \cite{ghlrewp}.

\item{The transition $\bskk$:}

As has been pointed out in \cite{bfewp}, the mixing-induced CP-violating
asymmetry $\acpmi(\bskk)$ may be used to determine the amplitude
$z\equiv T'/|P'|$ provided the CKM-angle $\gamma$ is known. On the other
hand, if one follows the approach to extract these quantities presented
above, the quantity $\xi^{(s)}_{K^+K^-}$ containing essentially all the
information needed to describe the CP-violating effects arising in $\bskk$
(see ref.~\cite{bfewp}) can be predicted and allows an interesting test.
Although the branching ratio BR$(B_s\to K^+K^-)$ is rather promising
and expected to be of $\order(10^{-5})$, the large $B^0_s-\bar B^0_s-$mixing
parameter may cause experimental problems.

\item{The transition $B_d\to\pi^0K_{\mbox{{\scriptsize S}}}$:}

In the notation of Gronau et al.~\cite{ghlrewp}, the transition amplitude
of the decay $B^0_d\to\pi^0K^0$ takes the form
\bea
\lefteqn{A(B^0_d\to\pi^0K^0)=-\frac{1}{\sqrt{2}}\left[C'-P'+(c_u-c_d)\pewp
-c_d\pewpc\right]\approx\,\,\,\mbox{}}\nonumber\\
&&\frac{1}{\sqrt{2}}\left[P'-(c_u-c_d)\pewp\right]
\approx A(\bar B^0_d\to\pi^0\bar K^0),\lab{e17}
\eea
where we have performed the same approximations as in \cite{bfewp}, i.e.\
have neglected the colour-suppressed tree-level and electroweak penguin
contributions which are both $\order(\bar\lambda^2)$. Within this
approximation, mixing-induced and direct CP violation in
$B_d\to\pi^0K_{\mbox{{\scriptsize S}}}$ (the final state
$|\pi^0K_{\mbox{{\scriptsize S}}}\rangle$ is an eigenstate of the
${\cal CP}$ operator) are characterized by
\bea
\acpmi(B_d\to\pi^0K_{\mbox{{\scriptsize S}}})&=&-\sin2\beta\lab{e18}\\
\acpdir(B_d\to\pi^0K_{\mbox{{\scriptsize S}}})&=&0.\lab{e19}
\eea
An exact definition of these asymmetries can be found in
\cite{rf8}. Sizable non-vanishing direct CP violation would indicate
that the approximation of neglecting the $C'$, $\bar C'$ topologies,
which has also been performed in the approach \cite{bfewp} to determine
$(c_u-c_d)\pewp$, is not very good. In this case, (\ref{e18})
does not allow a clean measurement
of the CKM-angle $\beta$. Comparing (\ref{e18}) with the absolutely clean
asymmetry $\acpmi(B_d\to\psi K_{\mbox{{\scriptsize S}}})=-\sin2\beta$,
one can also obtain information on the quality of this approximation and
moreover on the amplitudes $C'$, $\bar C'$. Note that the branching ratio
BR$(B_d^0\to\pi^0 K_{\mbox{{\scriptsize S}}})=\mbox{BR}(B_d^0\to\pi^0K^0)/2$
can also be predicted with the help of the results of ref.~\cite{bfewp}.

\item{The substructure of the $B(\bar B)\to\pi\pi$ isospin triangles:}

Combining the analyses of refs.~\cite{bfewp,bfalpha}, we are in a position
to resolve the substructure of the $B\to\pi\pi$ isospin triangle as has
been shown in Fig.~3. In drawing this figure, we have neglected terms
of $\order(\bar\lambda^3)$, e.g.\ colour-suppressed
$\bbtodb$ electroweak penguins.
A similar construction can also be performed for the corresponding
CP-conjugate modes. In particular the dashed triangle, which is related
to $C$, $T$ and $C+T$, can be fixed. The theoretical accuracy of the
determination of the amplitudes $C$ and $T$ is limited by the unknown
$\order(\bar\lambda^2)$ exchange and $SU(3)$-breaking topologies $E$ and
$P_3$, respectively \cite{ghlrsu3}. Taking into account \cite{ghlrewp}
\beq\lab{e20}
A(B^0_s\to\pi^0\bar K^0)=-\frac{1}{\sqrt{2}}\left[C-P+(c_u-c_d)\pew\right],
\eeq
where colour-suppressed $\bbtodb$
electroweak penguins have been neglected as in
Fig.~3, a prediction of the direct and mixing-induced CP-violating
asymmetries arising in the mode $B_s\to\pi^0 K_{\mbox{{\scriptsize S}}}$
is possible and would allow another consistency check if it should become
possible to measure these quantities in future experiments, maybe in those
of the $3^{\mbox{{\scriptsize rd}}}$ generation.

\end{itemize}

In summary, using the $SU(3)$ flavour symmetry of strong interactions,
we have presented strategies for extracting the quantities $z$, $\bar z$
and the CKM-angle $\gamma$ which are needed as the central input for the
approach to control electroweak penguins in nonleptonic $B$-decays
presented in ref.~\cite{bfewp}. In particular an approximate method making
use of the modes $\bppipko$, $\bdpimkp$, $\bdbpipkm$ and $B^+\to\pi^+\pi^0$
should be rather promising from the experimental point of view. We have
derived a transparent analytical expression allowing a {\it quantitative}
test of the question whether the ratio ${\cal R}_{\mbox{{\scriptsize EW}}}
\equiv|(c_u-c_d)\pewp|/|T'|$ determined by following the approach
\cite{bfewp} is in accordance with its description through the Standard
Model of electroweak interactions. As a by-product, this
formula yields an upper bound $|\Delta\alpha|\leqsim 6^\circ$ for the
uncertainty originating from electroweak penguins in the
$\alpha$-determination by means of $B(\bar B)\to\pi\pi$ decays
\cite{bfewp,gl}. Some interesting predictions and consistency checks
involving the decays $B_s\to\pi^0(\eta,\Phi)$, $B_s\to K^+ K^-$ and
$B_d\to\pi^0 K_{\mbox{{\scriptsize S}}}$ have also been discussed.
Moreover, we have pointed out that the substructure of the
$B(\bar B)\to\pi\pi$ isospin triangles fixing e.g.\ the CP-violating
asymmetries arising in the mode $B_s\to\pi^0K_{\mbox{{\scriptsize S}}}$
can be resolved in principle. The strategies and results presented in this
letter in combination with those of refs.~\cite{bfewp,bfalpha} should allow
valuable insights into the world of the electroweak penguins and should
furthermore provide an interesting test of the Standard Model.

\vspace{0.5cm}

I would like to thank Andrzej Buras for a very enjoyable collaboration on
topics related to this paper.

\vspace{1cm}

\section*{Figure Captions}
\begin{table}[htb]
\begin{tabular}{ll}
Fig.\ 1:&A determination of the quantities $z$, $\bar z$ and the CKM-angle
$\gamma$.\\
&\\
Fig.\ 2:&The dependence of ${\cal R}_{\mbox{{\scriptsize EW}}}
\equiv|(c_u-c_d)\pewp|/|T'|$ on the top-quark mass\\
&$m_t$ for $a_1=1$, $r_{SU(3)}=1$ and various values of the
CKM-parameter $R_b$.\\
&\\
Fig.\ 3:&The substructure of the $B\to\pi\pi$ isospin triangle.\\
&\\
\end{tabular}
\end{table}

\newpage

\begin{figure}[p]
\vspace{15cm}
\caption{}\lab{f1}
\end{figure}

\begin{figure}[p]
\vspace{15cm}
\caption{}\lab{f2}
\end{figure}

\begin{figure}[p]
\vspace{15cm}
\caption{}\lab{f3}
\end{figure}


\newpage

\begin{thebibliography}{99}
\bibitem{ck}L.L. Chau and W.-Y. Keung, \prl~{\bf 53} (1984) 1802.
\bibitem{js}C. Jarlskog and R. Stora, \pl~{\bf B208} (1988) 268.
\bibitem{gw}M. Gronau and D. Wyler, \pl~{\bf B265} (1991) 172.
\bibitem{dun}I. Dunietz, \pl~{\bf B270} (1991) 75.
\bibitem{adk}R. Aleksan, I. Dunietz and B. Kayser, \zp~{\bf C54}
(1992) 653.
\bibitem{kay}R. Aleksan, B. Kayser and D. London, National Science
Foundation preprint {\bf NSF-PT-93-4}, {\bf hep-ph/9312338} (1993).
\bibitem{grl}M. Gronau, J.L. Rosner and D. London, \prl~{\bf 73} (1994)
21.
\bibitem{ghlr}M. Gronau, O.F. Hern\'andez, D. London and J.L. Rosner,
\pr~{\bf D50} (1994) 4529.
\bibitem{hlgr}O.F. Hern\'andez, D. London, M. Gronau and J.L. Rosner,
\pl~{\bf B333} (1994) 500.
\bibitem{hl}O.F. Hern\'andez and D. London, Universit\'e de Montr\'eal
preprint {\bf UdeM-LPN-TH-94-198}, {\bf hep-ph/9406418} (1994).
\bibitem{lon}D. London, Universit\'e de Montr\'eal preprint {\bf
UdeM-LPN-TH-94-199}, {\bf hep-ph/9406412} (1994).
\bibitem{ghlrsu3}M. Gronau, O.F. Hern\'andez, D. London and J.L. Rosner,
Technion preprint {\bf TECHNION-PH-95-10}, {\bf hep-ph/9504326} (1995).
\bibitem{rfewp1}R. Fleischer, \zp~{\bf C62} (1994) 81.
\bibitem{rfewp2}R. Fleischer, \pl~{\bf B321} (1994) 259.
\bibitem{rfewp3}R. Fleischer, \pl~{\bf B332} (1994) 419.
\bibitem{dhewp1}N.G. Deshpande and X.-G. He, \pl~{\bf B336} (1994) 471.
\bibitem{dht}N.G. Deshpande, X.-G. He and J. Trampetic,
\pl~{\bf B345} (1995) 547.
\bibitem{dy}D. Du and M. Yang, Institute of High Energy Physics (Beijing)
preprint {\bf BIHEP-TH-95-8}, {\bf hep-ph/9503278} (1995).
\bibitem{dhewp2}N.G. Deshpande and X.-G. He, \prl~{\bf 74} (1995) 26.
\bibitem{ghlrewp}M. Gronau, O.F. Hern\'andez, D. London and J.L. Rosner,
Technion preprint {\bf TECHNION-PH-95-11}, {\bf hep-ph/9504327} (1995).
\bibitem{dhgam}N.G. Deshpande and X.-G. He, University of Oregon preprint
{\bf OITS-576}, {\bf hep-ph/9505369} (1995).
\bibitem{bfewp}A.J. Buras and R. Fleischer, Universit\"at Karlsruhe preprint
{\bf TTP95-29}, {\bf hep-ph/9507303} (1995).
\bibitem{bf}A.J. Buras and R. Fleischer, \pl~{\bf B341} (1995) 379.
\bibitem{gl}M. Gronau and D. London, \prl~{\bf 65} (1990) 3381.
\bibitem{sm}S.L. Glashow, \np~{\bf 22} (1961) 579; S. Weinberg,
\prl~{\bf 19} (1967) 1264; A. Salam, in ``Elementary Particle Theory'',
edited by N. Svartholm (Almqvist and Wiksell, Stockholm, 1968).
\bibitem{wolf}L. Wolfenstein, \prl~{\bf 51} (1983) 1945.
\bibitem{bfalpha}A.J. Buras and R. Fleischer, Universit\"at Karlsruhe preprint
{\bf TTP95-30}, {\bf hep-ph/9507460} (1995).
\bibitem{kp}G. Kramer and W.F. Palmer, {\bf DESY 95-131},
{\bf hep-ph/9507329} (1995).
\bibitem{bjl}A.J. Buras, M. Jamin and M.E. Lautenbacher,
\np~{\bf B408} (1993) 209.
\bibitem{rf}R. Fleischer, \zp~{\bf C58} (1993) 483.
\bibitem{kps}G. Kramer, W.F. Palmer and H. Simma, \np~{\bf B428}
(1994) 77.
\bibitem{bb}G. Buchalla and A.J. Buras, \np~{\bf B398} (1993) 285;
\np~{\bf B400} (1993) 225.
\bibitem{kam}A.N. Kamal, Int.\ J.\ Mod.\ Phys.\ {\bf A7} (1992) 3515.
\bibitem{bgr}A.J. Buras, J.-M. G\'erard and R. R\"uckl, \np~{\bf B268}
(1986) 16.
\bibitem{bsw}M. Bauer, B. Stech and M. Wirbel, \zp~{\bf C29} (1985) 637
and \zp~{\bf C34} (1987) 103.
\bibitem{blo}A.J. Buras, M.E. Lautenbacher and G. Ostermaier, \pr~{\bf D50}
(1994) 3433.
\bibitem{al}A. Ali and D. London, {\bf DESY 95-148},
{\bf hep-ph/9508272} (1995).
\bibitem{nrsx}M. Neubert, V. Rieckert, B. Stech and Q.P. Xu, in ``Heavy
Flavours'', edited by A.J. Buras and M. Lindner (World Scientific,
Singapore, 1992).
\bibitem{cleo}M.S. Alam et al., CLEO Collaboration, \pr~{\bf D50} (1994) 43.
\bibitem{bur}A.J. Buras, \np~{\bf B434} (1995) 606.
\bibitem{il}T. Inami and C.S. Lim, Progr.\ Theor.\ Phys.\ {\bf 65}
(1981) 297 and Progr.\ Theor.\ Phys.\ {\bf 65} (1981) 1772.
\bibitem{rf8}R. Fleischer, \pl~{\bf B341} (1994) 205.

\end{thebibliography}
\end{document}